\documentclass[11pt, letterpaper]{article}

\usepackage{enumitem}
\usepackage{color}
\usepackage{graphicx, amssymb, amsmath, fullpage, dsfont, amsthm, braket}
\usepackage{authblk,soul}
\usepackage[utf8]{inputenc}
\usepackage[english]{babel}
\usepackage[x11names]{xcolor} 
\usepackage[colorlinks,
            linkcolor = blue,
            urlcolor  = blue,
            citecolor = blue,
            anchorcolor = blue]{hyperref}
\usepackage{tikz}
\usetikzlibrary{calc}

\newcommand{\splitter}[5]{\pic at (#1) {splitter};
\coordinate (#2) at (A);\coordinate (#3) at (B);\coordinate (#4) at (C);\coordinate (#5) at (D)}

\newcommand{\collector}[1]{\draw [->] (#1) -- +(0.2,0) pic[right=1ex]{collector}}

\newcommand{\shifter}[4]{\pic[#3] at (#1) {shifter};
\coordinate (#2) at (b);\node at (a) {#4}}

\tikzset{pics/.cd,
collector/.style={code={\draw[fill=red!20] (0,0.25) arc(90:-90:0.5cm and 0.25cm) -- cycle;}},shifter/.style={code={\draw[fill=gray!20]  (#1+90:0.3) rectangle coordinate[midway] (a) +(-0.8,-0.6);
\draw[-] (#1+180:0.8) -- +(-0.4,0) coordinate (b);}},
splitter/.style={code={\draw[ultra thick,blue] (#1:{sqrt(1/2)}) --
(#1+180:{sqrt(1/2)});
\draw[-] (#1+45:{sqrt(0.5)}) -- (#1+225:{sqrt(0.5)});
\draw[-] (#1+45:{sqrt(0.5)}) -- +(0.4,0) coordinate (C);
\draw[-] (#1+225:{sqrt(0.5)}) -- +(-0.4,0) coordinate (B);
\draw[-] (#1-45:{sqrt(0.5)}) -- (#1+135:{sqrt(0.5)});
\draw[-] (#1-45:{sqrt(0.5)}) -- +(0.4,0) coordinate (D);
\draw[-] (#1+135:{sqrt(0.5)}) -- +(-0.4,0) coordinate (A);
}},splitter/.default=0}

\usetikzlibrary{patterns,decorations}
\usetikzlibrary{angles, quotes}
\usepackage{pgfplots,caption}
\pgfplotsset{compat=1.17}

\usepackage{fullpage}
\usepackage{doi}

\DeclareMathOperator{\tr}{Tr}

\numberwithin{equation}{section}

\newcommand{\be}{\begin{equation}}
\newcommand{\ee}{\end{equation}}

\begin{document}

\title{\textbf{Superresolution at the quantum limit beyond two point sources}}
\author{Hari Krovi}
\affil[]{Quantum Engineering and Computing\\Raytheon BBN Technologies, Cambridge, MA}

\maketitle

\begin{abstract}
Superresolution refers to the estimation of parameters of an image with an accuracy beyond standard classical techniques such as direct detection. In seminal work by Lu et al., \cite{tsang2016quantum}, a measurement to estimate the separation distance of two point sources (with a known centroid) was shown to achieve the quantum Cramer-Rao bound. This work made implicit use of reflection symmetry of the sources. Here we present a framework that uses more general symmetry in a constellation to construct a quantum measurement that achieves the quantum Cramer-Rao bound in estimation of parameters. We show how this technique can be used to estimate parameters simultaneously in symmetric point-source constellations with more than two point sources. In order to use symmetry explicitly, we make use discrete point spread functions in momentum space that maintain this symmetry. This framework allows us to use techniques from quantum computing such as Fourier transforms and linear optical circuits to implement the optimal measurement. To our knowledge, this is first work that shows for more than two point sources achievable quantum limits of estimation and modal transformations.
\end{abstract}

\section{Introduction}
For the problem of estimation of an unknown parameter, techniques from quantum parameter estimation can help achieve a mean squared error (MSE) scaling better than the one achieved using direct detection followed by classical signal processing. In seminal work by Lu, Nair and Tsang \cite{tsang2016quantum}, it was shown that estimating the distance between two point sources can be done better with modal transformations before detection than by direct detection of the photons followed by signal processing. In fact, the MSE achieved using modal transformations was independent of the separation distance whereas the MSE achieved by direct detection scaled inversely with the separation distance. The fact that the MSE of classical techniques diverges as the separation distance goes to zero is called Rayleigh's curse and \cite{tsang2016quantum} showed that modal transformations in the optical domain can lift this curse. In general, the estimation of any parameter with a mean squared error that scales better than direct detection is usually termed \emph{superresolution}.

In subsequent work, this was extended to other scenarios such as two point sources off-axis \cite{sliver}. Other works considered more general sources such as thermal states \cite{PhysRevLett.117.190801} and arbitrary quantum states in \cite{PhysRevLett.117.190802}. The recent review by Tsang \cite{Tsang_review} gives a good overview of the topic and discusses related work in quantum imaging. The work of Lupo et al in \cite{lupo2020quantum} show that linear interferometry can achieve the quantum limit in the paraxial regime and moreover, considered a model of the imaging system that gives rise to a discrete state space i.e., it is essentially a qudit state space. The point spread functions we consider are based on this and give rise to very similar states. In \cite{Discrete_quantum_imaging}, this is called \emph{discrete quantum imaging} and it is shown that calculations of the quantum Fisher information can be simplified in this model. However, as the number of parameters increases, it becomes hard to compute the optimal measurement that achieves the quantum limit. In fact, in \cite{zhou2019modern} it was shown that at most two moments can be independent of the object's size for arbitrary objects. This put bounds on the number of parameters that can be \emph{superresolved} for arbitrary sources. The works that considered multiple point source localization only compared their schemes to direct detection and other metrics since there seems to be no meaningful way to define quantum limits for multiple parameters that cannot be estimated simultaneously. Here we show that multiple point sources with symmetry with at most two unknown parameters can be estimated at the quantum limit generalizing SPADE and SLIVER. As far as we are aware, there has been no work showing that the quantum limit can be attained for more than two point sources.

Quantum imaging refers to modal transformations on the light field before detection followed by classical signal processing. At the mathematical level, the difference between quantum and classical imaging is that in quantum imaging, one can modify the probability distribution one gets after detection by performing quantum operations on the light field, whereas in direct detection, one has fixed distributions. To modify these distributions, one needs to implement mode transformations on the light field. To this end, we introduce techniques from quantum computing such as abelian Fourier transforms to quantum imaging that help achieve superresolution. These transforms are discrete and typically act on a finite set of modes. We describe next how the symmetry of the constellation can be maintained by choosing the right point spread function.

The main idea is to engineer a point spread function (psf) that respects the symmetry to the constellation and construct modal transformations that achieve the quantum limit in the paraxial regime. We find that choosing a psf similar to the one in \cite{lupo2020quantum} allows us to use the symmetry of the constellation in two dimensions. The idea that symmetry is useful was already observed by Lu et al \cite{tsang2016quantum} where they consider a reflection symmetric psf and in \cite{sliver}, the authors notice that two dimensional symmetry was useful in the construction of the measurement. In \cite{lupo2020quantum}, with a different detector model, it was observed that the symmetry of the two element abelian group ($\mathbb{Z}_2$) and the Fourier transform can be used to estimate separation of two point sources at the quantum limit. In \cite{Paur2018tempering}, the symmetric and anti-symmetric subspaces are used to construct the quantum optimal measurement for distance estimation of two point sources. However, these papers do not use symmetry to go beyond two point sources. Here we consider multiple point sources including an $N$-point source ring, which can be generalized to an extended source (of an annular ring, whose radius is to be estimated). Our framework can be applied to parameter estimation for any symmetric quantum state, not just one that comes from an imaging system.

Several papers go beyond sources with symmetry and consider parameter estimation problems for arbitrary source locations and differing brightness. While the schemes presented perform better than direct detection, they have not been shown to attain the quantum limit of estimation. In fact, in \cite{Projective_measurements}, the authors showed that estimating the centroid, separation and the unequal brightness of even two point sources cannot be done simultaneously. In \cite{PhysRevLett.126.120502}, the authors show that adding noise to the imaging system adversely affects superresolution. In \cite{PhysRevA.104.022410}, a scheme to estimate the centroid of a linear array constellation of points is given. Extended sources were also considered in some papers. In \cite{PhysRevA.99.033847}, it was shown numerically that estimating the length of a line (rather than the distance between two points) can also be done using a Hermite-Gauss mode sorter.

This paper is organized as follows. In section \ref{sec:estimation_theory}, we discuss the mathematics of quantum estimation theory that we will use and how group symmetry can be exploited. In section \ref{sec:model}, we define the model of the sources and the imaging system that we use and in section \ref{sec:examples}, we discuss how the theory from section \ref{sec:estimation_theory} can be used for different point source constellations to find parameters at the quantum limit. Finally, in section \ref{sec:conclusions}, we discuss the conclusions from this work and some open questions.

\section{Quantum estimation of states with symmetry}\label{sec:estimation_theory}
\subsection{Preliminaries on quantum estimation theory}
The theory of quantum Fisher information in estimation theory was developed in Helstrom \cite{Helstrom}, Nagaoka \cite{Nagaoka}, Braunstein and Caves \cite{BC94}. We give a brief description of this theory. We consider estimation of parameters $\Theta=\{\theta_1\dots \theta_k\}$ from a quantum state $\rho$ when there is a symmetry group associated with it i.e., when $\rho$ commutes with the operators of a symmetry group. First, we briefly recall the basics of quantum estimation theory. For a given quantum state $\rho$ parameterized by a set of parameters $\Theta$, the quantum Fisher information matrix (QFIM) $\mathcal{F}$ is defined as follows.
\be
\mathcal{F}_{\mu,\nu}=\text{Re}(\tr(\mathcal{L}_\mu\mathcal{L}_\nu\rho))\,,
\ee
where $\mu$ and $\nu$ belong the set $\Theta$ and $\mathcal{L}_\mu$ is the symmetric logarithmic derivative (SLD), which is defined as the solution of 
\be
\frac{d\rho}{d\theta_\mu}=\frac{1}{2}(\mathcal{L}_\mu\rho+\rho\mathcal{L}_\mu)
\ee
If the density matrix $\rho$ has the spectral decomposition given by eigenvalues $\lambda_i$ and eigenvectors $\ket{e_i}$, then the SLD can be written as 
\be
\mathcal{L}_\mu=\sum_{m,n:\lambda_m+\lambda_n\neq 0}\frac{2}{\lambda_m+\lambda_n}\bra{e_m}\frac{\partial\rho}{\partial\theta_\mu}\ket{e_n}\ket{e_m}\bra{e_n}\,.
\ee

In general, the matrix $\partial \rho/\partial \theta_\mu$ can have support outside the span of $\ket{e_i}$. However, if the eigenvectors are independent of the parameters (as is the case when there is symmetry), then the support of the derivative is still the span of $\ket{e_i}$. In this case, we see that
\be
\frac{\partial\rho}{\partial\theta_\mu}=\sum_i \frac{\partial\lambda_i}{\partial\theta_\mu}\ket{e_i}\bra{e_i}\,.
\ee
Plugging this back into the SLD, we get
\be
\mathcal{L}_\mu=\sum_{n}\frac{1}{\lambda_n}\frac{\partial\lambda_n}{\partial\theta_\mu}\ket{e_n}\bra{e_n}\,.
\ee
This means that the QFIM becomes
\be
\mathcal{F}_{\mu,\nu}=\sum_n\frac{1}{\lambda_n}\frac{\partial\lambda_n}{\partial\theta_\mu}\frac{\partial\lambda_n}{\partial\theta_\nu}\,.
\ee
This is exactly the classical Fisher information of the probability distribution $\lambda$. Since the classical CRB is achievable, the above shows that when the eigenvectors of $\rho$ are independent of the parameters to be estimated, then one can achieve the quantum Cramer-Rao bound by measuring in the eigenbasis of $\rho$. We will see next that when $\rho$ has a so called \emph{multiplicity-free} representation of a symmetry group associated with it, then it is possible to measure in the eigenbasis of $\rho$. We focus on abelian groups below for simplicity but this is true for non-abelian groups as well.

\subsection{Abelian group symmetry}
In quantum imaging of coherent point sources, the density matrix comes from an ensemble of pure states each corresponding to a specific point source. Suppose that the states that comprise the density matrix have an abelian group symmetry i.e., there is a unitary $U_g$ for every $g$ in the group $G$ such that $\ket{\psi_g}=U_g\ket{\psi_e}$, where $\ket{\psi_e}$ is some state associated with the identity element (this can be any state in the ensemble). The density matrix is given by
\be
\rho=\frac{1}{|G|}\sum_{g\in G}\ket{\psi_g}\bra{\psi_g}\,.
\ee
Such a state can be diagonalized using the quantum Fourier transform over the group $G$. The (unnormalized) eigenstates of $\rho$ turn out to be
\be
\ket{e_\lambda}=\frac{1}{\sqrt{|G|}}\sum_g \chi_\lambda(g) \ket{\psi_g}\,,
\ee
where $\chi_\lambda(g)$ is an irreducible character evaluated at $g$. First, we show that these states are orthogonal, even though $\ket{\psi_g}$ may not be orthogonal. To see this, consider the inner product between two such states.
\be
\braket{e_\lambda | e_{\lambda^\prime}} = \frac{1}{|G|}\sum_{g,g^\prime} \chi_\lambda(g^{-1})\chi_{\lambda^\prime}(g^\prime) \braket{\psi_g|\psi_{g^\prime}}\,.
\ee
Using the group translation property, we can re-write it as
\be
\braket{e_\lambda | e_{\lambda^\prime}} = \frac{1}{|G|}\sum_{g,g^\prime} \chi_\lambda(g^{-1})\chi_{\lambda^\prime}(g^\prime) \braket{\psi_e|U_{g^{-1}g^\prime}|\psi_e}\,.
\ee
Using orthogonality properties of characters and taking $g^{-1}g^\prime=g^{\prime\prime}$, this sum can be shown to be
\be
\braket{e_\lambda | e_{\lambda^\prime}} = \delta_{\lambda,\lambda^\prime}\sum_{g^{\prime\prime}} \chi_\lambda(g^{\prime\prime})\braket{\psi_e|U_{g^{\prime\prime}}|\psi_e}\,.
\ee
As can be seen above, the states are orthogonal and the norm of each depends on the inner product structure of the states $\ket{\psi_g}$. Now to see that $\rho$ is diagonalized with these states, first let us write $\ket{\psi_g}$ in terms of $\ket{e_\lambda}$. This can done as follows.
\be
\ket{\psi_g}=\frac{1}{\sqrt{|G|}}\sum_\lambda \chi^\ast_\lambda(g)\ket{e_\lambda}\,.
\ee
Using this, we can write the density matrix as
\be
\rho=\frac{1}{|G|^2}\sum_{g,\lambda,\lambda^\prime}\chi^\ast_\lambda(g)\chi^\ast_{\lambda^\prime}(g^{-1})\ket{e_\lambda}\bra{e_{\lambda^\prime}}=\frac{1}{|G|}\sum_\lambda \ket{e_\lambda}\bra{e_{\lambda}}\,,
\ee
where we have used the orthogonality properties of characters again. The eigenvalues of $\rho$ are $n_\lambda/|G|$, where $n_\lambda=\braket{e_\lambda | e_\lambda}$. We will use these concepts later in Section~\ref{sec:examples} to show how to construct the quantum-optimal measurement for symmetric sources.

\section{Model of the source and the imaging system}\label{sec:model}
We will work in a setting where there are $N$ sources of equal brightness that are all in the $XY$ plane relative to the imaging system. We assume that the point sources are coherent sources of light. In order to be able to describe a linear optical circuit to implement the quantum-optimal measurement using beamsplitters and phaseshifters, we also assume that the there is at most one photon in the modes collected. This is true when the sources are in the far-field so that one can apply the paraxial approximation (this approximation was also made in \cite{tsang2016quantum} and subsequent work).

Next, we define the model for the imaging system that we use in our calculations, namely, the point spread function that we pick and the modal transformation needed before detecting the photons. Since we assume that the modes contain at most one photon, we only need single photon detectors. The point spread function that we pick is a collection of delta functions in Fourier (or momentum) space. The locations of these delta functions is given by the symmetry of the sources. For instance, if the sources are two points along the $X$ axis (with an unknown separation distance $d$ that we want to estimate), then there are two delta functions in momentum space at $p_1$ and $p_2$, which are separated by a known distance $p$ (which can be differed from $d$). In general, the delta functions will be momenta in $2$d momentum space at locations $p_i^x,p_i^y$ where these $XY$ locations have the same symmetry as the point sources, but with known but possibly different values than the locations of the point sources.

This choice of psf is equivalent to the model of the imaging system defined in \cite{lupo2020quantum}. In that model, the imaging system has $N_C$ collectors, where each collector is assumed to be a pin-hole that can couple to one spatial mode of light. The collectors are assumed to be in the $XY$ plane. The positions of the collectors are assumed to be $\mathbf{c}_i=(u_i,v_i)$ for $i=1,\dots N_C$.  In the paraxial approximation, at most one photon arrives in a given temporal mode at the image plane. The quantum state generated due to $N_S$ sources (all assumed to be in the XY plane) in locations $\mathbf{r}_i=(x_i,y_i)$ is as follows.
\be\label{eq:density_matrix}
\rho = \frac{1}{\sqrt{N_S}}\sum_{i=1}^{N_S} \ket{\psi_i}\bra{\psi_i}\,,
\ee
where
\be
\ket{\psi_i}=\frac{1}{\sqrt{N_C}}\sum_{j=1}^{N_C} \exp(-\frac{ik}{z_0}\mathbf{r}_i\cdot\mathbf{c}_j)\ket{j}\,,
\ee
where $z_0$ is the distance between the object plane and the aperture plane.

This model can be derived from the one above with the delta function psf. The quantum state of a single source (in the paraxial regime where at most a single photon is obtained in a time slot) is given by
\be
\ket{\psi}=\int d\mathbf{s} \,\psi(\mathbf{s}-\mathbf{r})\ket{\mathbf{s}}\,,
\ee
where $\mathbf{r}$ is the position of the source and $\psi(\mathbf{s})$ is the point spread function of the imaging system. Looking at this in the momentum space, we have
\be
\ket{\tilde{\psi}}=\int d\mathbf{p}\,\exp(-i\mathbf{p}\cdot\mathbf{r})\tilde{\psi}(\mathbf{p})\ket{\mathbf{p}}\,.
\ee
Now if we choose the Fourier transformed psf $\tilde{\psi}(\mathbf{p})$ to be delta functions at specific points $\mathbf{c}_i$ in the momentum space, then it can be seen that this gives the model described above after normalizing by the number of points i.e.,
\begin{equation}
    \ket{\tilde{\psi}}=\sum_j \frac{1}{\sqrt{N}}\exp(-i\mathbf{p}_j\cdot\mathbf{r})\ket{j}\,.
\end{equation}

Choosing the psf to be as above means that we can mathematically model the resulting quantum states in terms of qudits. This allows us to use qudit techniques to diagonalize density matrices with symmetry. Specifically, we will see that the quantum Fourier transform is the modal transformation that achieves the quantum limit (the quantum Cramer-Rao bound). In addition, in the paraxial regime, since we can assume that there is a single photon in any qudit basis state, we can use the techniques of \cite{PhysRevLett.73.58} to decompose the modal transformation into a circuit composed of beamsplitters and phaseshifters.

\section{Constellations with symmetry}\label{sec:examples}
In this section, we consider the problem of estimating parameters of constellations of sources with symmetry and discuss a comparison to direction detection. Then we present a number of examples with multiple sources and parameters.
\subsection{Modal transformation to achieve the quantum limit}
Suppose that the sources have a symmetry group $G$ that preserves the density matrix. The action of this group can be thought of as a permutation of the sources. For example, for a constellation of $N$ sources with circular symmetry, the group will be $\mathbb{Z}/N\mathbb{Z}$ of integers modulo $N$. This group is a subgroup of the orthogonal group which preserves inner products. The unitary representation of this group on the space of momenta is given by permuting the momentum states according to the same group element. In order to be able to do this, we need to assume that the number of momenta is the same as the number of sources and that they are arranged with the same symmetry (e.g., circular) as the sources in the momentum plane. In this situation, if $g$ is any group element that rotates or reflects the sources and takes the source labeled $i^\prime$ to the source labeled $i$, then we have 
\begin{align}
\ket{\psi_i}&=\frac{1}{\sqrt{N}}\sum_{j=1}^{N} \exp(-i\mathbf{p}_{j}\cdot g\mathbf{r}_{i^\prime})\ket{j}\\
&=\frac{1}{\sqrt{N}}\sum_{j=1}^{N} \exp(-i g\mathbf{p}_{j}\cdot \mathbf{r}_{i^\prime})\ket{j}\,.
\end{align}
This can be written as
\begin{align}
&=\frac{1}{\sqrt{N}}\sum_{j=1}^{N} \exp(-i\mathbf{p}_{j}\cdot\mathbf{r}_{i^\prime})U_{g}\ket{j}\\
&=U_{g}\ket{\psi_{i^\prime}}\,.
\end{align}

This means that the group $G$ is a group of symmetries of the density matrix $\rho$ via the unitary representation $U$. If this representation is multiplicity-free i.e., it contains each Fourier component exactly once, then we can completely diagonalize the density matrix. To be more concrete, we give below some examples with various types of planar symmetry.

We note that the above formalism works even if the number of momentum points is greater than the number of sources. We choose them to be equal for simplicity of analysis. If the number of delta functions in the psf are greater than the number of sources, then we need to have all the momenta arranged so that the symmetry of the sources is embedded in them. For instance, if the sources have circular symmetry, then at least $N$ momenta in the psf must be arranged in a circle.

The transformation to be performed before detection to diagonalize the density matrix is given by the quantum Fourier transform. The unitary matrix that corresponds to the quantum Fourier transform has rows labeled by $\lambda$ and columns labeled by $g$. The matrix entries are given by
\begin{equation}
    [U_{QFT}]_{\lambda,g}=\frac{1}{\sqrt{|G|}}\chi_\lambda(g^{-1})\,.
\end{equation}
As mentioned above, in the paraxial regime, we can assume that the qudit basis has only one photon per basis state. This means that the above transformation is a modal transformation and can be decomposed into a circuit with beamsplitters and phaseshifters \cite{PhysRevLett.73.58}. In the examples below, we show how this decomposition works. Before we move to the examples, let us examine the performance of direct detection on a constellation with this psf. 

In the paraxial regime, direct detection of the modes amounts to measuring in the basis of the momenta (labeled $\ket{j}$ above). It can be seen from the states above that if we measure in this basis, the probability of getting any particular basis state is $1/N$. This distribution contains no information about the unknown parameters. The Fisher information of estimating any unknown parameter with this distribution is zero since the distribution is independent of the parameters. This psf illustrates the stark difference between quantum techniques and classical direct detection. Other psfs fare better with direct detection. Indeed it is known that for separation estimation of two point sources \cite{tsang2016quantum}, with the Gaussian psf, direct detection has a performance that matches the quantum Fisher information asymptotically as the separation distance increases (but goes to zero as the separation distance decreases). This points to how psf engineering is an important additional degree of freedom to be considered in addition to modal transformations to get to the quantum limit of parameter estimation.

\subsection{1-D reflection symmetry with two point sources}
We start with the problem that has already been well studied in several papers starting with the seminal paper \cite{tsang2016quantum}. With the discrete psf, it was also done in \cite{lupo2020quantum}. This example will orient the reader to the formalism described above. In this problem, we assume that the centroid of the constellation is known and that the sources are at a distance $r$ from the centroid and we are interested in estimating $r$. The quantum state received at the aperture is given by 
\be
\rho=\frac{1}{2}(\ket{\psi_1}\bra{\psi_1} + \ket{\psi_2}\bra{\psi_2})\,,
\ee
where
\begin{align}
&\ket{\psi_1}=\frac{1}{\sqrt{2}}(\exp(-ipr)\ket{0} + \exp(ipr))\ket{1})\\
&\ket{\psi_2}=\frac{1}{\sqrt{2}}(\exp(ipr)\ket{0} + \exp(-ipr))\ket{1})\,,
\end{align}
where we assume that the psf is chosen with delta functions that are located at $p$ and $-p$ from the origin on the $X$ axis in the image plane. The eigenvalues of the density matrix turn out to be as follows.
\be
\lambda_1 =\frac{1}{2}( 1+\Re(\braket{\psi_1|\psi_2})), \, \lambda_2 = \frac{1}{2}(1-\Re(\braket{\psi_1|\psi_2}))\,,
\ee
where $\Re(\cdot)$ stands for real part. Since 
\be
\braket{\psi_1|\psi_2}=\cos(2p r)\,,
\ee
we have that the eigenvalues are
\be
\lambda_1 =\cos^2(pr), \, \lambda_2 = \sin^2(pr)\,.
\ee
The QFI can be calculated to be
\be
\text{QFI}=4 p^2\,.
\ee

\begin{figure}
\begin{center}
\begin{tikzpicture}
\draw [->,thick] (0,-2) -- (0,2) node (yaxis) [left] {$y$};
\draw [->,thick] (-2,0) -- (2,0) node (xaxis) [right] {$x$};
\draw [arrows=<->,thick, Azure4] (-1,0.5) -- (1,0.5) ;
\node at (-0.3,0.7) {$2r_0$};
\foreach \i in {0,180}{%
\filldraw [blue]  (\i:1cm) circle (1.5pt);}

\end{tikzpicture}
\end{center}
\caption{Two point sources on-axis.}
\end{figure}
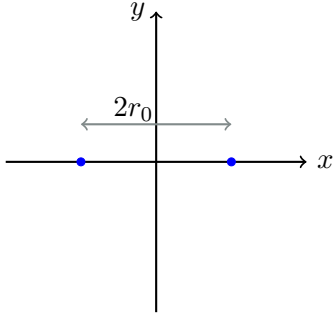

\begin{figure}
\begin{center}
\begin{tikzpicture}[>=stealth]
\coordinate (s1) at (0,0);
\splitter{s1}{s1a}{s1b}{s1c}{s1d};

\node [left] at (s1a) {$m_1$};
\node [left] at (s1b) {$m_2$};
\collector{s1c};
\collector{s1d};

\end{tikzpicture}
\end{center}
\caption{Linear optical circuit to estimate the separation distance between two point sources on the $X$ axis. The modes $m_1$ and $m_2$ are mixed on a $50:50$ beamsplitter.}\label{fig:TwoPointSourcesCircuit}
\end{figure}
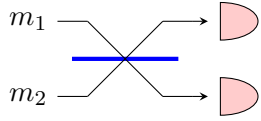

The linear optical circuit to perform the QFI achieving measurement is given in Fig. \ref{fig:TwoPointSourcesCircuit}. The simple linear optical circuit above achieves superresolution in the estimation of separation distance when the probability distribution obtained at the detector is used for further signal processing that achieves the classical Cramer-Rao bound.

\subsection{1D off axis}
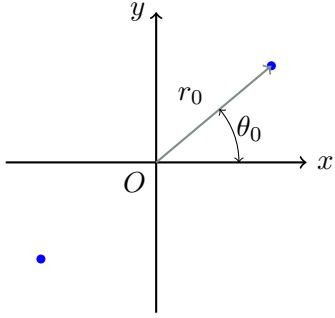
\begin{figure}
\begin{center}
\begin{tikzpicture}
\draw [->,thick] (0,-2) -- (0,2) node (y) [left] {$y$};
\draw [->,thick] (-2,0) -- (2,0) node (x) [right] {$x$};


\coordinate[label=below left:$ O $] (O) at (0,0);
\foreach \i in {40,220}{%
\filldraw [blue]  (\i:2cm) circle (1.5pt);}
\draw[thick,Azure4,->] (0, 0) -- (40:2);
\coordinate[label=above left:$ r_0 $] (z) at (40:1);
\pic [draw, <->,
      angle radius=11mm, angle eccentricity=1.2,
      "$\theta_0$"] {angle = x--O--z};

\end{tikzpicture}
\end{center}
\caption{Two point sources off-axis.}\label{fig:two point sources off axis}
\end{figure}

Next, we consider two point sources off-axis as shown in Fig. \ref{fig:two point sources off axis}. The two points are at locations $(x,y)$ and $(-x,-y)$. Here the goal is to estimate the two parameters $x$ and $y$ or equivalently, the radius $r$ and the angle $\theta$. The symmetry in this case is still independent of the parameters. We pick the delta functions in momentum space at $p$ and $-p$ on the $X$ axis as in the previous example. The density matrix is given by 
\begin{equation}
    \rho=\frac{1}{2}(\ket{\psi_1}\bra{\psi_1}+\ket{\psi_2}\bra{\psi_2})\,,
\end{equation}
where 
\begin{align}
    &\ket{\psi_1}=\frac{1}{\sqrt{2}}(\exp(irp\cos\theta)\ket{0} + \exp(-irp\cos\theta)\ket{1})\,,\\
    &\ket{\psi_2}=\frac{1}{\sqrt{2}}(\exp(-irp\cos\theta)\ket{0} + \exp(irp\cos\theta)\ket{1})\,.
\end{align}
This shows that interchanging the states $\ket{0}$ and $\ket{1}$ is a symmetry of the density matrix since it interchanges $\ket{\psi_1}$ and $\ket{\psi_2}$.

This symmetry implies that the eigenvectors of the density matrix are \begin{align}
    &\ket{+}=\frac{1}{\sqrt{2}}(\ket{0}+\ket{1})\\
    &\ket{-}=\frac{1}{\sqrt{2}}(\ket{0}-\ket{1})\,.
\end{align}
The corresponding eigenvalues turn out to be
\begin{align}
    \lambda_+=\cos^2(rp\cos\theta)\\
    \lambda_-=\sin^2(rp\cos\theta)\,.
\end{align}
We can calculate the quantum Fisher information of estimating the separation distance using the eigenvalues. This turns out to be
\begin{equation}
    QFI=4p^2\cos^2(\theta)\,,
\end{equation}
showing that the QFI is independent of the separation distance. One can think of the angle $\theta$ as being misalignment. This shows that for any fixed misalignment $\theta$, superresolution still holds and the quantum limit is independent of the separation distance. If instead of picking the psf as delta functions at $p$ and $-p$, we had picked them to be at radius $p$ and angle $\theta_0$, then the QFI of estimating separation would become
\begin{equation}
    QFI=4p^2\cos^2(\theta-\theta_0)\,.
\end{equation}
Therefore, an initial guess (based on direct detection using some of the photons) can increase the superresolution constant and bring it closer to $4p^2$.

\subsection{2D reflection symmetry with four point sources}
\begin{figure}
\begin{center}
\begin{tikzpicture}
\draw [->,thick] (0,-2) -- (0,2) node (yaxis) [left] {$y$};
\draw [->,thick] (-2,0) -- (2,0) node (xaxis) [right] {$x$};
\draw [arrows=<->,thick, Azure4] (-0.95,0.5) -- (0.95,0.5) ;
\node at (-0.3,0.7) {$2x$};

\draw [arrows=<->,thick, Azure4] (1.1,0.45) -- (1.1,-0.45) ;
\node at (1.35,0.2) {$2y$};

\foreach \i in {22,158,202,-22}{%
\filldraw [blue]  (\i:1cm) circle (1.5pt);}

\end{tikzpicture}
\end{center}
\caption{Four point sources}\label{fig:2D-4sources}
\end{figure}
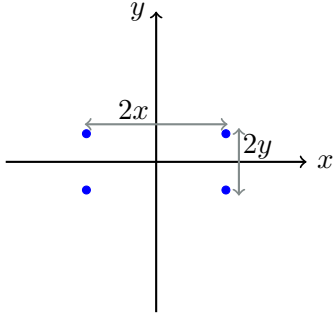

The next example we consider is a constellation with four point sources located at the corners of a rectangle in the XY plane, where the sides of the rectangle are $2x_0$ and $2y_0$ assumed unknown (see Fig. \ref{fig:2D-4sources}). The centroid is assumed to be known and is assumed to be the origin. The density matrix is an equal mixture of four states given by 
\be
\ket{\psi_i}=\sum_j \exp(-i\mathbf{p}_j\cdot \mathbf{r}_i)\ket{j}\,.
\ee
Here $\mathbf{r}_i=(x_i,y_i)$ are the positions of the sources. If we assume that the momentum space delta functions are $\mathbf{p}_j=(x_j^\prime,y_j^\prime)$, then we can write the above states as
\be
\ket{\psi_i}=\sum_j \exp(-i(x_i x^\prime_j + y_iy_j^\prime))\ket{j_x,j_y}=\ket{\psi_i^{(x)}}\otimes\ket{\psi_i^{(y)}}\,,
\ee
where
\be
\ket{\psi_i^{(x)}}=\sum_{j_x} \exp(-ix_i x^\prime_j)\ket{j_x}\,,
\ee
and similarly for $\ket{\psi_i^{(y)}}$. The indices $j_x$ and $j_y$ run over two terms and $\ket{j_x}$ and $\ket{j_y}$ can be taken to be qubits for the purpose of calculating the QFI. The optimal measurements are the ones that diagonalize each tensor copy separately, which reduces this to a tensor product of two 1-D cases. The QFI matrix has diagonal entries $4x^2$ and $4y^2$.

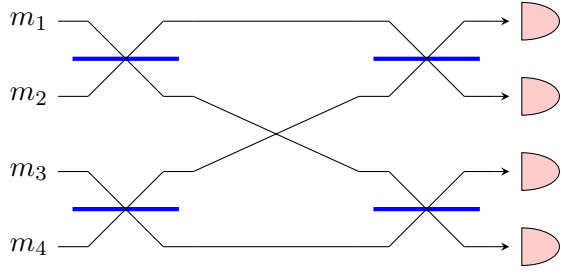
\begin{figure}
\begin{center}
\begin{tikzpicture}[>=stealth]
\coordinate (s1) at (-2,1);
\coordinate (s2) at (-2,-1);
\coordinate (s3) at (2,1);
\coordinate (s4) at (2,-1);

\splitter{s1}{s1a}{s1b}{s1c}{s1d};
\node [left] at (s1a) {$m_1$};
\node [left] at (s1b) {$m_2$};

\splitter{s2}{s2a}{s2b}{s2c}{s2d};
\node [left] at (s2a) {$m_3$};
\node [left] at (s2b) {$m_4$};

\splitter{s3}{s3a}{s3b}{s3c}{s3d};

\splitter{s4}{s4a}{s4b}{s4c}{s4d};
\draw[-] (s2d) -- (s4b) (s1d) -- (s4a) (s1c) -- (s3a) (s2c) -- (s3b);

\collector{s3c};
\collector{s3d};
\collector{s4c};
\collector{s4d};

\end{tikzpicture}
\end{center}
\caption{Linear optical circuit to estimate the separation distances between four point sources. The input modes are labeled $m_1$ through $m_4$ and the beamsplitters are all $50:50$ beamsplitters.}\label{fig:FourPointSourcesCircuit}
\end{figure}

The linear optical circuit to diagonalize the density matrix and the SLD operators is in Fig. (\ref{fig:FourPointSourcesCircuit}). This circuit is essentially the Hadamard or Walsh transform for a unary encoded linear optical circuit. This is the quantum Fourier transform over the abelian group $\mathbb{Z}_2^{2}$.

\subsection{N point sources with circular symmetry}
\begin{figure}
\begin{center}
\begin{tikzpicture}
\draw [->,thick] (0,-3.5) -- (0,3.5) node (yaxis) [left] {$y$};
 \draw [->,thick]  (-3.5,0) -- (3.5,0) node (xaxis) [right] {$x$};
\foreach \i in {0,22.5,45,...,360}{%
\filldraw [blue]  (\i:3cm) circle (1.5pt);}
\draw[thick,Azure4,->] (0, 0) -- (45:3);
\node at (1,0.8){$r$};
\end{tikzpicture}
\end{center}
\caption{$N$ point sources}
\end{figure}
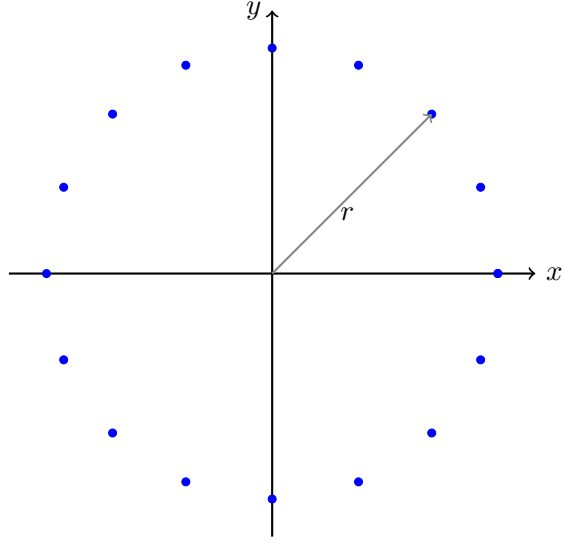

We now consider the problem of estimating the radial distance $r_0$ of $N$ sources arranged in a circle around a known centroid. The angular separation $\theta_0$ between two adjacent sources in the object plane is also assumed to be known. The sources are assumed to have the same brightness. This is a different generalization of the two point source problem. This problem has a circular symmetry that can be exploited. The density matrix can be written as above in Eq.~\ref{eq:density_matrix}. The states $\ket{\psi_i}$ would inherit this circular symmetry. In this case, cyclic shift of the momentum states $j$ would generate all the states $\ket{\psi_i}$ from one of them. To be more precise, suppose $U$ is a cyclic shift of the states taking $j$ to $j+1$ modulo $N$ as follows.
\be
U\ket{j}=\ket{j+1 \text{ mod } N}\,.
\ee
Such a shift operator acting on a source state $\ket{\psi_i}$ would give the following. We assume that addition of indices is modulo $N$ and $R(\theta_0)$ is counter-clockwise rotation of the co-ordinates.
\begin{align}
U\ket{\psi_i}&=\frac{1}{\sqrt{N}}\sum_{j=1}^{N} \exp(-\mathbf{r}_i\cdot\mathbf{p}_j)\ket{j+1}\\
&=\frac{1}{\sqrt{N}}\sum_{j=1}^{N} \exp(-\mathbf{r}_i\cdot R(\theta_0)\mathbf{p}_j)\ket{j}\\
&=\frac{1}{\sqrt{N}}\sum_{j=1}^{N} \exp(-R(\theta_0) \mathbf{r}_i\cdot \mathbf{p}_j)\ket{j}\\
&=\ket{\psi_{i+1}}\,.
\end{align}

This symmetry implies that we can diagonalize the density matrix using the Fourier basis. Therefore, measuring in the Fourier basis would achieve the quantum Fisher information. The QFI can be calculated using the eigenvalues of the density matrix, which are as follows
\be
\lambda_\ell=\frac{1}{N}\sum_{m=0}^{N-1}\exp(2\pi im\ell/N)\braket{\psi_m|\psi_0}=\frac{1}{N^2}\sum_{m,n=0}^{N-1}\exp(2\pi im\ell/N)\exp(-(\mathbf{r}_m-\mathbf{r}_0)\cdot\mathbf{p}_n)\,.
\ee
These can be re-written as follows.
\begin{equation}
    \lambda_k=\Bigg|\frac{1}{N}\sum_{n=0}^{N-1}\exp(2\pi i nk/N)\exp(-ip_0r_0\cos(2\pi n/N))\Bigg|^2\,.
\end{equation}
Writing $\lambda_k=|a_k|^2$, we can write the derivative of $\lambda_k$ with respect to $r_0$ (the unknown radius) as
\begin{equation}
    \lambda_k^\prime=a_k (a_k^\ast)^\prime + a_k^\prime a_k^\ast\,.
\end{equation}
Notice that $a_k$ is just the inverse discrete Fourier transform of the following function. 
\begin{equation}
    f(n)=\exp(-ipr\cos(n\theta))\,,
\end{equation} 
where $\theta=2\pi/N$. The derivative of $a_k$ is
\begin{equation}
    a_k^\prime=\frac{1}{N}\sum_{n=0}^{N-1}\exp(ink\theta)\exp(-ipr\cos(n\theta))(-ip\cos(n\theta))\,,
\end{equation}
which is the inverse Fourier transform of the function $f^\prime(n)$. Now the QFI is given by
\begin{equation}
    QFI=\sum_k\frac{1}{\lambda_k}(\lambda_k^\prime)^2=\sum_k\frac{4}{\lambda_k}|a_k|^2|a_k^\prime|^2=\sum_k 4|a_k^\prime|^2\,.
\end{equation}
By Parseval's theorem, the QFI becomes
\begin{equation}
    QFI=\sum_n \frac{4}{N} |f^\prime(n)|^2=\frac{4p^2}{N}\sum_n \cos^2(n\theta)\,.
\end{equation}
This can be evaluated to be
\begin{equation}
    QFI=2p^2\sum_n \frac{1}{N}(1+\cos(2n\theta))=2p^2\sum_n\frac{1}{N}\bigg(1+\frac{1}{2}\bigg(\exp(2in\theta)+\exp(-2in\theta)\bigg)\bigg)\,.
\end{equation}
Now we have that the sum over each exponential is
\begin{equation}
    \sum_n \frac{1}{N}\exp(\pm 2in\theta) = 
    \begin{cases} 1 &\mbox{if } N = 2 \\
0 & \mbox{if } N>2\,. \end{cases}
\end{equation}
Therefore, the QFI of estimating $r$ turns out to be
\begin{equation}
    QFI=\begin{cases} 4p^2 &\mbox{if } N = 2 \\
2p^2 & \mbox{if } N>2\,. \end{cases}
\end{equation}

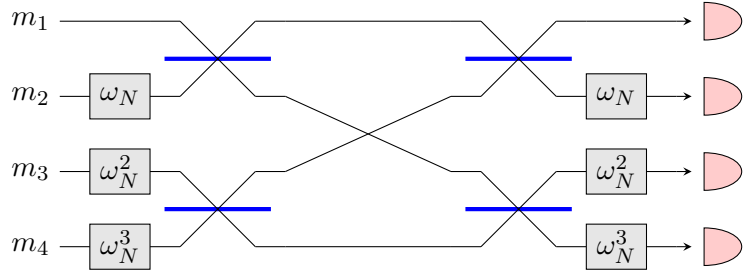
\begin{figure}
\begin{center}
\begin{tikzpicture}[>=stealth]
\coordinate (s1) at (-2,1);
\coordinate (s2) at (-2,-1);
\coordinate (s3) at (2,1);
\coordinate (s4) at (2,-1);

\splitter{s1}{s1a}{s1b}{s1c}{s1d};
\splitter{s2}{s2a}{s2b}{s2c}{s2d};
\splitter{s3}{s3a}{s3b}{s3c}{s3d};
\splitter{s4}{s4a}{s4b}{s4c}{s4d};

\shifter{s1b}{sh1}{rotate=0}{$\omega_N$};
\shifter{s2a}{sh2}{rotate=0}{$\omega^2_N$};
\shifter{s2b}{sh3}{rotate=0}{$\omega^3_N$};
\draw[-] (s1a) -- (s1a-|sh1) node[left] {$m_1$};
\node [left] at (sh1) {$m_2$};
\node [left] at (sh2) {$m_3$};
\node [left] at (sh3) {$m_4$};

\shifter{s3d}{sh4}{rotate=180}{$\omega_N$};
\shifter{s4c}{sh5}{rotate=180}{$\omega^2_N$};
\shifter{s4d}{sh6}{rotate=180}{$\omega^3_N$};

\draw[-] (s2d) -- (s4b);
\draw[-] (s1d) -- (s4a);
\draw[-] (s1c) -- (s3a);
\draw[-] (s2c) -- (s3b);

\draw[-] (s3c) -- (s3c-|sh4) coordinate (sh7);
\collector{sh7};
\collector{sh4};
\collector{sh5};
\collector{sh6};

\end{tikzpicture}
\end{center}
\caption{Linear optical circuit to estimate the radius of a circularly symmetric configuration of $N$ point sources (here $N=4$). The input modes are labeled $m_1$ through $m_4$ and the beamsplitters are all $50:50$ beamsplitters. The square boxes are phaseshifters with the phase labeled inside (here $\omega_N=e^{2\pi i/N}$).}\label{fig:NPointSourcesCircuit}
\end{figure}

The circuit to perform this measurement can be constructed based on the Fourier coefficients. As a network of beamsplitters and phaseshifters, it is shown in Fig. (\ref{fig:NPointSourcesCircuit}) for four point sources in a circle.

\section{Conclusions}\label{sec:conclusions}
We have presented a scheme to use symmetry of a constellation to estimate parameters of interest at the quantum limit. In the literature so far, calculations of the quantum limit were limited to constellations two point sources, where it can be shown that several measurements can achieve the quantum Cramer-Rao bound. In our examples, we have presented constellations with an arbitrary number of sources. For instance, the number of points in our constellations can be arbitrarily large for circular symmetry. This allows one approximate an annular disc--an extended object and estimate its radius beyond the Rayleigh limit.

We have also presented structured linear optical circuits that allow one to implement these measurements. Linear optical circuits are very useful for certain implementations of quantum computing such as the Knill-Laflamme-Milburn scheme \cite{KLM} (and its successors). This allows for the application of programmable arrays of linear optical elements for imaging at the quantum limit.

It would be interesting to see if the work here can be extended to three-dimensions. The psf we have used only allow for two-dimensional imaging but recently certain types of psfs such as the double helix psf \cite{Pavani2995} have been shown to be capable of 3-D localization. Extending the results in this paper to the double helix psf would allow for parameters of more types of constellations to optimally estimated.

\section{Acknowledgments}
I am grateful to Jason Fleischer, Mike Gehm, Gordon Wetzstein, Saikat Guha, Amit Ashok, Luke Govia, Kwan Kit Lee, Michael Grace, Aqil Sajjad and Mark Neifeld for discussions about superresolution, point spread functions and quantum limits. This material is based on work supported by the Defense Advanced Research Projects Agency (DARPA) under Agreement No. HR00112090128. 

\bibliographystyle{ieeetr}
\bibliography{Quantum_imaging}
\end{document}